# Blind Restoration of Real-World Audio by 1D Operational GANs

Turker Ince, Serkan Kiranyaz, Ozer Can Devecioglu, Muhammad Salman Khan, Muhammad Chowdhury, and Moncef Gabbouj, *Fellow, IEEE*

*Abstract*— **Objective:** Despite numerous studies proposed for audio restoration in the literature, most of them focus on an isolated restoration problem such as denoising or dereverberation, ignoring other artifacts. Moreover, assuming a noisy or reverberant environment with limited number of fixed signal-to-distortion ratio (SDR) levels is a common practice. However, real-world audio is often corrupted by a *blend* of artifacts such as reverberation, sensor noise, and background audio mixture with *varying* types, severities, and duration. In this study, we propose a novel approach for blind restoration of real-world audio signals by Operational Generative Adversarial Networks (Op-GANs) with temporal and spectral objective metrics to enhance the quality of restored audio signal regardless of the type and severity of each artifact corrupting it. **Methods:** 1D Operational-GANs are used with *generative* neuron model optimized for blind restoration of any corrupted audio signal. **Results:** The proposed approach has been evaluated extensively over the benchmark TIMIT-RAR (speech) and GTZAN-RAR (non-speech) datasets corrupted with a random blend of artifacts each with a random severity to mimic real-world audio signals. Average SDR improvements of over 7.2 dB and 4.9 dB are achieved, respectively, which are substantial when compared with the baseline methods. **Significance:** This is a pioneer study in blind audio restoration with the unique capability of direct (time-domain) restoration of real-world audio whilst achieving an unprecedented level of performance for a wide SDR range and artifact types. **Conclusion:** 1D Op-GANs can achieve robust and computationally effective real-world audio restoration with significantly improved performance. The source codes and the generated real-world audio datasets are shared publicly with the research community in a dedicated GitHub repository[1].

## I. INTRODUCTION

AUDIO signals in real-world environments are usually corrupted by different types of background mixture(s), interfering speech, sensorial noise, and possibly room reverberation (e.g., see Figure 1, part-a). Any arbitrary combination of these artifacts with varying severities makes the restoration very challenging since no prior assumption can be made. Yet such a blind restoration is a crucial need for many audio processing and communication applications such as automatic speech recognition (ASR), speaker identification, voice calls, or hearing-assistive devices. Ideally, they require a robust restoration performance against any type of audio artifact (e.g., see Figure 1, part-b) with high computational efficiency.

The problem of audio restoration when recorded from a single-channel microphone, especially under nonstationary noise, reverberation, and a mixture of other audio sources has been an active research area [1]. In the past, researchers have developed many DSP-oriented and model-based techniques [1]-[5]. However, most of them have been specifically designed for the sole purpose of denoising or dereverberation of audio clips, assuming only one or a few degradation sources with pre-specified SDR levels exist. Despite this fact, their restoration performance for real-world audio signals is usually limited.

Recently, Deep Learning (DL) approaches utilizing the variants of Generative Adversarial Networks (GANs) or Deep Neural Networks (DNNs) have been employed either for audio restoration or spectral mapping, or even mask prediction. They have achieved significant performance improvements over traditional methods [6]. The first GAN application was the speech enhancement GAN (SEGAN) [7], which is an end-to-end framework based on a 22-layer 1D convolutional neural network (CNN) in an encoder-decoder architecture operating directly on the 1D raw audio. A secondary L1 norm factor is added to the loss of generator (G) to minimize the distance between the generated and clean samples. For a relatively small test set based on 5 *fixed* noise types at 4 *fixed* SDR levels (17.5, 12.5, 7.5, and 2.5 dB), SEGAN achieved some improvements in 5 different speech quality metrics and surpassed the Wiener filter by 2.66 dB segmental signal-to-noise-ratio (SSNR). Following works based on SEGAN, such as ISEGAN and DSEGAN [8], tried to tackle the challenges of raw audio waveform-based networks by employing multiple chained generators but achieved insignificant gains in performance.

Other researchers applied 2D GANs on spectral domain representations of the audio such as Short-Time Fourier Transform (STFT) and log-Mel filterbank spectra. In [9],

T. Ince, is with the Electrical and Electronics Engineering Department, Izmir University of Economics, Izmir, Turkey (email:turker.ince@ieu.edu.tr).

S. Kiranyaz, M. S. Khan, M. Chowdhury are with the Electrical Engineering Department, Qatar University, Doha, Qatar (e-mails: mkiranyaz@qu.edu.qa, salman@qu.edu.qa, mchowdhury@qu.edu.qa ).

O. Devecioglu and M. Gabbouj are with the Department of Computing Science, Tampere University, Tampere, Finland (e-mail: moncef.gabbouj@tuni.fi , ozer.devecioglu@tuni.fi ).

FSEGAN was proposed to perform spectral feature mapping based on the log-Mel filterbank features. They applied their model as input to an existing automatic speech recognition (ASR) model and improved the word error rate (WER) of the existing system by %7.

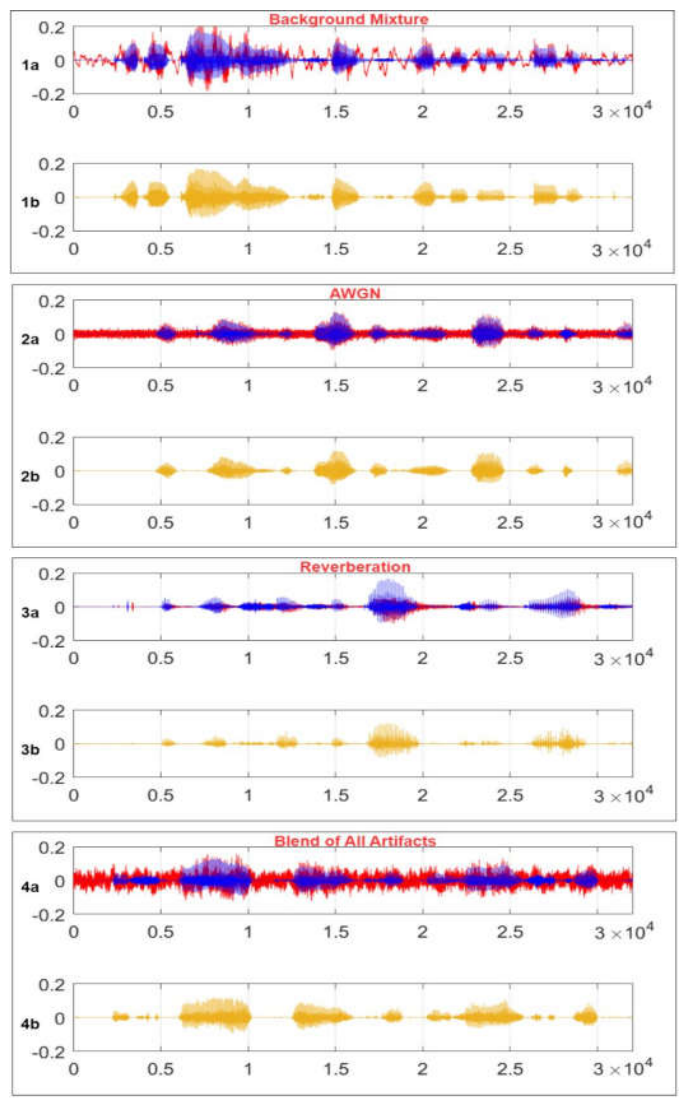

Figure 1: A sample set of 2-second segments from the TIMIT dataset signals corrupted by different artifacts and the corresponding restored signals using the proposed method. Original and corrupted audio are plotted in blue and red (part a), and the restored audio is plotted in yellow (part b).

HiFi-GAN was introduced in [10], based on a feed-forward WaveNet together with multi-scale adversarial training in both the time domain and time-frequency domain for denoising and dereverberation of audio signals. They utilize three waveform discriminators and another discriminator on the Mel-spectrogram with adversarial training to enhance the perceptual quality of the augmented speech corrupted only with the other audio sources at random SDR in the range of 10-30 dB. Several studies have focused on directly optimizing the human-auditory perception-related performance metrics in their models, such as QualityNet [11] and MetricGAN [12]. In [11], an approximated perceptual evaluation of speech quality (PESQ) metric is estimated using a supervised end-to-end CNN-based network, called Quality-Net, from input magnitude spectrograms. Slightly higher PESQ scores than the corrupted counterparts are achieved over the test set which contains randomly selected 100 clean utterances from the TIMIT corpus [13] and corrupted only with one of the three audio mixtures (engine, street, and baby cry) and Additive Gaussian White Noise (AWGN) at five fixed SDR levels (-6, 0, 6, 12, and 18 dB). Similarly, MetricGAN trained BLSTM-based generators with magnitude spectrogram as inputs and used the PESQ or short-time objective intelligibility (STOI) metrics as a part of the adversarial loss instead of the traditional $L_1$ or $L_2$ losses. Though the speech signals are corrupted by a single artifact at a fixed SDR level, low to moderate improvements in PESQ and STOI scores were achieved over the test set which consists of randomly selected 100 utterances from the TIMIT corpus mixed with one of the four noise types at five fixed SNR levels (-12 dB, -6 dB, 0 dB, 6 dB, and 12 dB).

Recently, a cycle-consistent GAN (Cycle-GAN)-based approach was proposed for speech dereverberation. The performance of two Cycle-GAN models using unpaired and paired data was found to be similar according to the objective evaluation metrics and subjective evaluations [14]. Two generators in the Cycle-GAN are both the 22-layer 2D UNets (composed of an encoder and a decoder) with the real and imaginary parts of the STFT as the 2D input. The models were trained with noise- and mixture-free synthetically reverberated data from the Librivox dataset and tested with both the Librivox [15] and AMI [16] datasets whereas the latter is composed of real reverberant audio clips. While both models perform comparable in terms of frequency-weighted segmental SNR (FWSegSNR) and estimated reverberation time (T60), the SDR improvement was only 0.9 dB by the paired GAN model versus no improvement for the unpaired model over the AMI dataset.

There are certain limitations and drawbacks of the abovementioned models. Although promising results and performance improvements over the traditional methods have been shown, most of these works were tested for a single artifact type, i.e., only against the audio mixtures alone (using a few distinct audio sources as the mixture model) and at a pre-determined SDR case, or for noise-free dereverberation. On the contrary, real-world audio (both speech and non-speech) can be corrupted by one, some, or all the artifact types each with varying (random) severities. In particular, using only a few mixture models or reverberation types at specified SDR levels can render the application of the proposed solutions in practice and adversely affect the generalization performance of these systems. As a result, such methods will obviously fail to restore properly any real-world audio signal corrupted with a *blend of* artifacts, as typical samples shown in Fig 1. Additionally, there are other known shortcomings of the specific approaches. Methods based on spectra transformation or spectral mask prediction can produce artifacts due to the loss of phase information. On the other hand, waveform-based methods (i.e., SEGAN, ISEGAN, DSEGAN) are known to have more artifacts and distortions compared to spectral methods [10]. Moreover, most methods first transform the 1D raw speech signal to 2D using STFT and then utilize 2D deep networks, both of which significantly increase the computational complexity.

To address these drawbacks and limitations, in this study, we address this problem with a *blind* restoration approach thus avoiding any prior assumption over the artifact types and severities. We propose a novel 1D operational GAN (Op-GAN) network where both generator (G) and discriminator are compact Self-Organized Operational Neural Networks (Self-ONNs) [17]. Recently, Self-ONNs have been shown to outperform their predecessors, CNNs, in many regression and classification tasks [17]-[25] with an elegant computational efficiency. Thanks to the generative neuron model with superior learning capabilities over the convolutional neurons, this study once again demonstrates that the shallowest GAN network ever proposed with only 10 layers in the generator network and 6 layers in the discriminator network will be sufficient to achieve an unprecedented SDR improvement over the real-world audio test dataset. Furthermore, this study proposes a novel loss function for the generator with the spectral loss term merged with the waveform based $L_1$ norm. This enables the generator to further improve its output waveform through the composite loss function from both time and frequency domains. Additionally, certain network configuration modifications have been performed to further boost the restoration performance. The final generator model trained and validated for the "corrupted" to "clean" audio segment transformation can then be used for blind audio restoration.

In order to emulate the real-world audio signals with any blend of artifacts, we first created a dataset that consists of 2703 original 2-second audio scripts (at 16 kHz) from the TIMIT corpus corrupted by *randomly* selected artifacts with *random* weights. We used a rich set of audio artifacts such as 15 background mixture models, AWGN as the sensory noise, and 12 room reverberation models (generated by a set of binaural room impulse responses (BRIRs)). Such a "random" blend leads to corrupted audio clips at varying SDR levels from -6 to 6 dB. We did not target audio with high SDR levels on purpose because one of the objectives of this study is to address the restoration of severely corrupted audio. The performance of the proposed approach is evaluated by computing the objective evaluation metrics including SDR, STOI, and PESQ [11].

We can summarize the novel and significant contributions of this study as follows:

1- This is a pioneer study where audio signal restoration is addressed as a "blind" approach thus avoiding any prior assumption such as a certain artifact type and its severity.

2- This is the first study where 1D Operational-GANs are proposed for an end-to-end audio signal restoration system working directly on the raw audio in real-time.

3- A novel loss function is developed for the generator to improve the restoration performance using the information from both time and frequency domains.

4- Despite the earlier studies in the literature aiming only the speech enhancement, this study targets the restoration of any audio type, particularly those with severe corruption (i.e., SDR ranging from -6 dB to 6 dB).

5- To emulate the real-world audio signals with any blend of artifacts, two real-world audio datasets are created with both clean and corrupted audio clips and shared publicly[1] along with the source codes [26].

6- The proposed approach has been tested over the benchmark datasets and achieved an unprecedented level of average SDR restoration performance (>7.2 dB on speech and >4.9 dB on the non-speech datasets, respectively). Finally, the proposed Self-ONN based Op-GAN is the most compact network model ever proposed for audio restoration with the least computational complexity.

The rest of the paper is organized as follows: a brief outline of 1D Self-ONNs, conditional GAN model, and the proposed approach with the Operational-GANs are introduced in Section II. The creation of benchmark datasets and a detailed set of experimental results are presented in Section III. Finally, Section IV concludes the paper and suggests topics for future research.

## II. PROPOSED APPROACH

In this section, we first briefly summarize Self-ONNs and their main properties. Then, we introduce the proposed approach by 1-D Self Operational GANs for audio restoration.

### A. 1D Self-Organized Operational Neural Networks

In this section, we briefly introduce the main network characteristics of 1D Self-ONNs[2] with their test and train mode formulations. While conventional CNNs have linear convolution operation and ONNs have fixed (static) nodal operators [33]-[41], a Self-ONNs with *generative* neurons can have any arbitrary nodal function, $\Psi$, (including possibly standard types such as linear and harmonic functions) optimized for each kernel element and for each connection. Obviously, Self-ONNs have the potential to achieve greater operational diversity and flexibility, allowing any nodal operator function to be formed during the training process.

The kernel elements of each generative neuron of a Self-ONN perform any nonlinear transformation, $\psi$, the function of which can be expressed by the Taylor-series near the origin ($a = 0$),

$$\psi(x) = \sum_{n=0}^{\infty} \frac{\psi^{(n)}(0)}{n!} x^n \qquad (1)$$

The $Q^{th}$ order truncated approximation, formally known as the Taylor polynomial, takes the form of the following finite summation:

$$\psi(x)^{(Q)} = \sum_{n=0}^{Q} \frac{\psi^{(n)}(0)}{n!} x^n \qquad (2)$$

The above formulation can approximate any nonlinear function $\psi(x)$ near 0. When the activation function bounds the neuron's input feature maps in the vicinity of 0 (e.g., *tanh*), the formulation in Eq. (2) can be exploited to form a composite nodal operator where the power coefficients, $\frac{\psi^{(n)}(0)}{n!}$, can be the parameters of the network learned during training.

---

[1] The optimized PyTorch implementation of the proposed Op-GANs and the two benchmark datasets, TIMIT-RAR and GTZAN-RAR, are publicly shared in https://github.com/InceTurker/Blind-Restoration-of-Real-World-Audio-by-1D-Operational-GANs

[2] The optimized PyTorch implementation of 1D Self-ONNs is publicly shared in https://github.com/junaidmalik09/fastonn.

It was shown in [21], [22], and [29] that the nodal operator of the $k^{th}$ generative neuron in the $l^{th}$ layer can take the following general form:

$$\widetilde{\psi_k^l}\left(w_{ik}^{l(Q)}(r), y_i^{l-1}(m+r)\right)$$
$$= \sum_{q=1}^{Q} w_{ik}^{l(Q)}(r,q)\left(y_i^{l-1}(m+r)\right)^q \quad (3)$$

Let $\widetilde{x_{ik}^l} \in R^M$ be the contribution of the $i^{th}$ neuron at the $(l-1)^{th}$ layer to the input map of the $l^{th}$ layer. Therefore, it can be expressed as,

$$\widetilde{x_{ik}^l}(m) = \sum_{r=0}^{K-1}\sum_{q=1}^{Q} w_{ik}^{l(Q)}(r,q)\left(y_i^{l-1}(m+r)\right)^q \quad (4)$$

where $y_i^{l-1} \in R^M$ is the output map of the $i^{th}$ neuron at the $(l-1)^{th}$ layer, $w_{ik}^{l(Q)}$ is a learnable kernel of the network, which is a $K \times Q$ matrix, i.e., $w_{ik}^{l(Q)} \in R^{K \times Q}$, formed as, $w_{ik}^{l(Q)}(r) = [w_{ik}^{l(Q)}(r,1), w_{ik}^{l(Q)}(r,2), \ldots, w_{ik}^{l(Q)}(Q)]$. By the commutativity of the summation operations in Eq. (4), one can alternatively write:

$$\widetilde{x_{ik}^l}(m) = \sum_{q=1}^{Q}\sum_{r=0}^{K-1} w_{ik}^{l(Q)}(r,q-1) y_i^{l-1}(m+r)^q \quad (5)$$

which can be simplified as:

$$\widetilde{x_{ik}^l} = \sum_{q=1}^{Q} Conv1D\left(w_{ik}^{l(Q)}, \left(y_i^{l-1}\right)^q\right) \quad (6)$$

Hence, the formulation can be accomplished by applying $Q$ 1D convolution operations. Finally, the output of this neuron can be formulated as follows:

$$x_k^l = b_k^l + \sum_{i=0}^{N_{l-1}} \widetilde{x_{ik}^l} \quad (7)$$

where $b_k^l$ is the bias associated with this neuron. The $0^{th}$ ($q = 0$) order term, the DC bias, is ignored as its additive effect can be compensated by the learnable bias parameter of the neuron. With the $Q = 1$ setting, a *generative* neuron reduces back to a convolutional neuron.

The efficient raw-vectorized formulations of the forward propagation, and detailed formulations of the Backpropagation (BP) training in a raw-vectorized form can be found in [27].

### B. 1D Operational-GANs

In the proposed work, we apply the extension of generative adversarial networks [28] to the conditional settings which were first introduced in [29] for transferring images from one representation to the other. In conditional GANs, the generator and discriminator are both dependent on some arbitrary external data (such as class labels or information from other modalities) and are capable of learning a multi-modal mapping from corrupted audio to clean audio through an adversarial learning process. In recent works [15]-[25], Self-ONNs outperformed conventional (deep) CNNs on various classification and segmentation tasks. To leverage this superiority in audio restoration, the proposed approach utilizes 1D Self-ONN layers in both generator and discriminator networks operating on raw audio waveform signals to enhance the learning capacity of a conditional GAN. The general framework of the proposed blind audio restoration scheme based on Operational-GAN (Op-GAN) is shown in Figure 2.

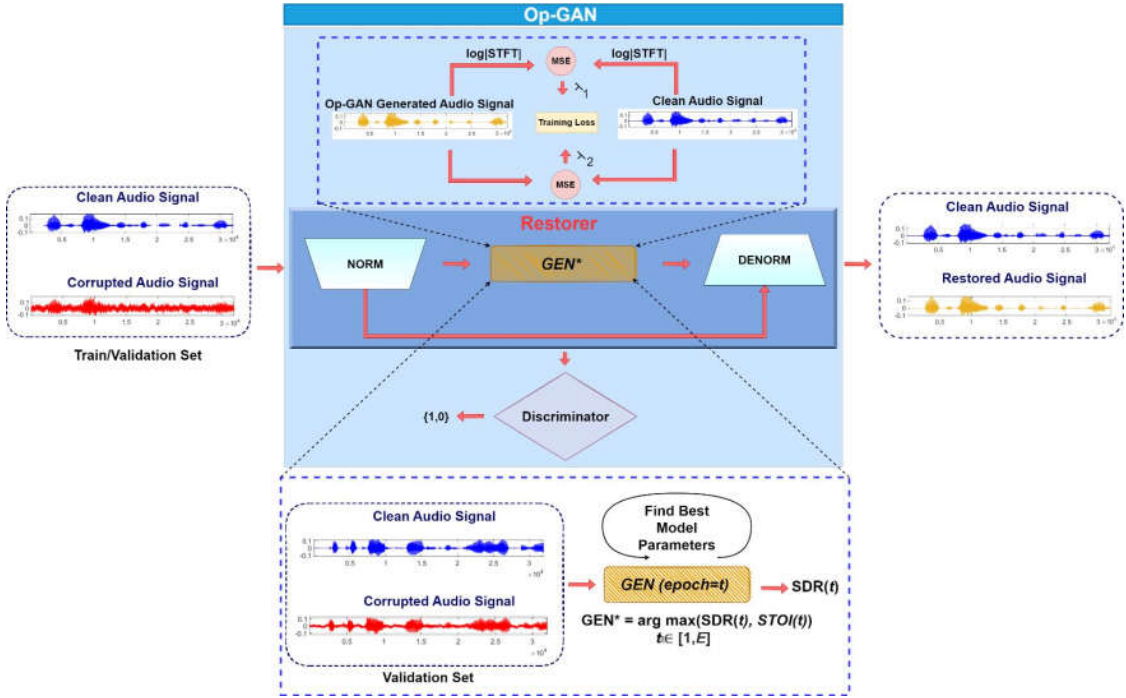

**Figure 2: The proposed blind audio restoration approach using 1D Operational-GANs.**

The proposed model applies time-domain segment-based restoration over input raw audio segments (of length $m = 32000$ samples corresponding to 2 seconds segment duration for the TIMIT dataset with a sampling rate of 16 kHz). To ensure an unbiased training on the type and severity of the corruption, each raw segment is corrupted with different (blend of) artifacts (e.g., 15 different background noise types, additive White Gaussian Noise (AWGN), and 12 reverberation models), each at varying severity levels controlled by a random weight. In brief, the ultimate objective is that the trained Op-GAN will learn to transform a "corrupted" segment to a "clean" segment regardless of, 1) its content, 2) artifact types, 3) artifact severities, and 4) audio volume level. The corrupted audio segments are generated to mimic the real-world audio by using the proposed scheme explained in Section III.A.

As shown in Figure 2, a raw audio segment from each batch is randomly selected as the input pair for the Op-GAN. First, volume-independent normalization is applied to each segment as follows:

$$X_N^s(i) = \frac{X^s(i)}{X_{max}^s} \quad (8)$$

where $X^s(i)$, $X_{max}^s$ and $X_N^s(i)$ are the original, maximum, and normalized pixel values for segment $s$, respectively. As in the conventional GANs, the generator network aims to learn the mapping of corrupted audio signals into clean counterparts. Through the proposed adversarial loss function, the generator can optimize its learning in both time and frequency domains. The discriminator helps the generator network to create more realistic clean audio signals by telling apart its synthesized signals from the real ones. Generally, the objective function of an Op-GAN can be expressed as [42],

$$\min_G \max_D L_{cGAN}(G, D) \\ = E[\log(D(X,Y))] \\ + E[\log(1 - D(X, G(X,z)))] \quad (9)$$

During the Back-Propagation training, the generator and discriminator play a two-player min-max game on the objective function $L_{cGAN}(G, D)$, where $G$ is the generator, $D$ is the discriminator, $z$ is the random noise and $Y$ is the corresponding output label. While the generator is aiming to minimize $\log(1 - D(X, G(X, z)))$, the discriminator's target is to maximize $D(X, Y)$ and $\log(1 - D(X, G(X, z)))$.

As suggested in [43], the cross-entropy loss in Eq. (9) is substituted by the least-squares loss functions to improve performance against the vanishing gradients problem:

$$\min_D L_{cGAN}(D) = \frac{1}{2} E_{X,Y \sim p_{data}(X,Y)}[(D(X,Y) - 1)^2] \\ + \frac{1}{2} E_{z \sim p_z(z)}[(D(G(X,z),Y)^2] \quad (10)$$

$$\min_G L_{cGAN}(G) \\ = \frac{1}{2} E_{z \sim p_z(z), Y \sim p_{data}(Y)}[((D(G(X,z),Y) - 1)^2] \quad (11)$$

In this work, to improve the audio restoration performance of Op-GAN, we propose to include both the waveform and spectral differences between the generated samples and the corresponding original samples based on the L2 norm (mean-squared error distance) to be added in (9) and (11). The $N$-point discrete Short-time Fourier Transform (STFT) of both the input and output signals are computed for spectral representation as in (12). For the window function, $W(n)$, $N$=256 samples long *Hanning* window with an overlap of 128 samples is used. These MSE loss metrics are controlled with the hyperparameters $\lambda_1$ and $\lambda_2$, which allow the generator to close the time and frequency domain gap between the fake and real audio samples and to better utilize the available time and frequency information of the audio samples for improved performance:

$$STFT[X, w, n] = X(n, w) = \sum_m X[m]W[n-m]e^{-jwm} \quad (12)$$

where the $N$-point discrete STFT, which is complex-valued, is obtained by sampling at each discrete radial frequency, $w = \frac{2\pi k}{N}$, and its magnitude is used in the spectral loss term:

$$Loss_{TD} = \|(Y(z) - G(X,z))\|_2 \quad (13)$$

$$Loss_{FD} = \|log_{10}(|STFT(Y(z)|) \\ - log_{10}(|STFT(G(X,z)|)\|_2 \quad (14)$$

The objective of Operational-GAN training is to minimize the total loss in (15):

$$Loss_{total} = L_{cGAN}(G) + \lambda_1 Loss_{TD} \\ + \lambda_2 Loss_{FD} \quad (15)$$

The weight parameters, $\lambda_1$ and $\lambda_2$, are used to balance the magnitudes of the corresponding temporal and spectral loss terms, and their values are determined experimentally (Section III.B). In the experiments, it is found that adversarial training works better together with the temporal and spectral losses computed at the generator output. The experimental setup and network parameters will be presented in the next section.

Additionally, we propose a modification in the Generator structure to replace the deconvolution (or transposed convolution) functions with upsampling and convolution operations (as shown in Figure 2) to avoid artifacts at the output of Op-GAN due to deconvolution. It is experimentally found that this approach further improves the audio restoration performance.

## III. EXPERIMENTAL RESULTS

In this section, we first introduce the generation of the two benchmark datasets to mimic real-world audio signals. Then, the experimental setup used for the evaluation of the proposed blind audio restoration approach will be presented. The comparative evaluations and the overall results of the experiments obtained using the train, validation, and test audio signals generated from the TIMIT (speech) and GTZAN (music) datasets will be presented next. Finally, both quantitative and qualitative evaluations will be performed. and the computational complexity of the proposed approach will be evaluated in detail.

## A. Real-World Audio Dataset Generation

Our aim in this study is to generate audio samples corrupted by a "blend of" artifacts with "random" severities, which is a commonality in a real-world acoustic environment. The proposed formation of both benchmark datasets generated in this study to mimic real-world corrupted audio clips is illustrated in Figure 3. To accomplish this aim, the outputs of randomly selected degradation sources are randomly (~U [0,1]) weighted before corrupting the clean target audio. The two artifacts (AWGN and background mixture) are *additive* while a linear convolution is applied for reverberation. Therefore, linear (random) weights can be used to control the weights of each artifact type selected. In addition to the random blend of all artifacts, we manually created single artifact cases (only one of $\alpha$, $\beta$ and $\gamma$ is turned on) to be included in the final datasets, which may correspond to the scenarios where only background mixture or reverberation exist.

For the evaluation of *speech* restoration, a total of 2703 clean data samples are taken from the TIMIT corpus [13], which contains recordings of different speakers from 8 major dialects of American English each reading phonetically rich sentences. Each utterance is a 2-second-long (32000 samples) segment with a sampling rate of 16 kHz. For the training and validation sets, 2000 randomly selected data samples are input to the real-world corrupted audio generation setup (Figure 3). The final train set includes 1500 samples from the blend of all artifacts as well as 500 samples per single artifact case, which adds up to a total of 3000 data samples. Note for each single artifact case samples are selected as non-overlapping groups (of 500 samples) from randomly selected 1500 train samples. Similarly, 500 and 703 randomly selected utterances from the remaining data are used to form the independent validation and test sets, which includes a total of 1000 and 1453 data samples, respectively. This benchmark dataset that can henceforth be used for real-world audio restoration is named TIMIT-RAR.

Similarly, for the evaluation of *non-speech* audio restoration, approximately 1.45-second-long segments (32000 samples with a sampling rate of 22050 Hz) from the classical and jazz music recordings of the GTZAN Music dataset are used. The final train set includes 1750 samples from the blend of all artifacts as well as 500 samples (as non-overlapping groups) per single artifact cases, which adds up to a total of 3250 data samples. Similarly, 500 and 830 randomly selected utterances from the remaining data are used to form the independent validation and test sets, which includes a total of 1000 and 1660 data samples, respectively. This benchmark dataset that can henceforth be used for real-world audio restoration is named GTZAN-RAR. The final train, validation, and test data compositions of both datasets are given in TABLE 1.

TABLE 1: TIMIT-RAR AND GTAN-RAR DATASETS COMPOSITIONS

|  |  | All Artifacts | AWGN | Mixture | Reverb. | Total |
|---|---|---|---|---|---|---|
| TIMIT-RAR | Train | 1500 | 500 | 500 | 500 | **3000** |
|  | Val. | 500 | 166 | 166 | 168 | **1000** |
|  | Test | 703 | 250 | 250 | 250 | **1453** |
| GTZAN-RAR | Train | 1250 | 500 | 500 | 500 | **3250** |
|  | Val. | 500 | 166 | 166 | 168 | **1000** |
|  | Test | 830 | 276 | 276 | 278 | **1660** |

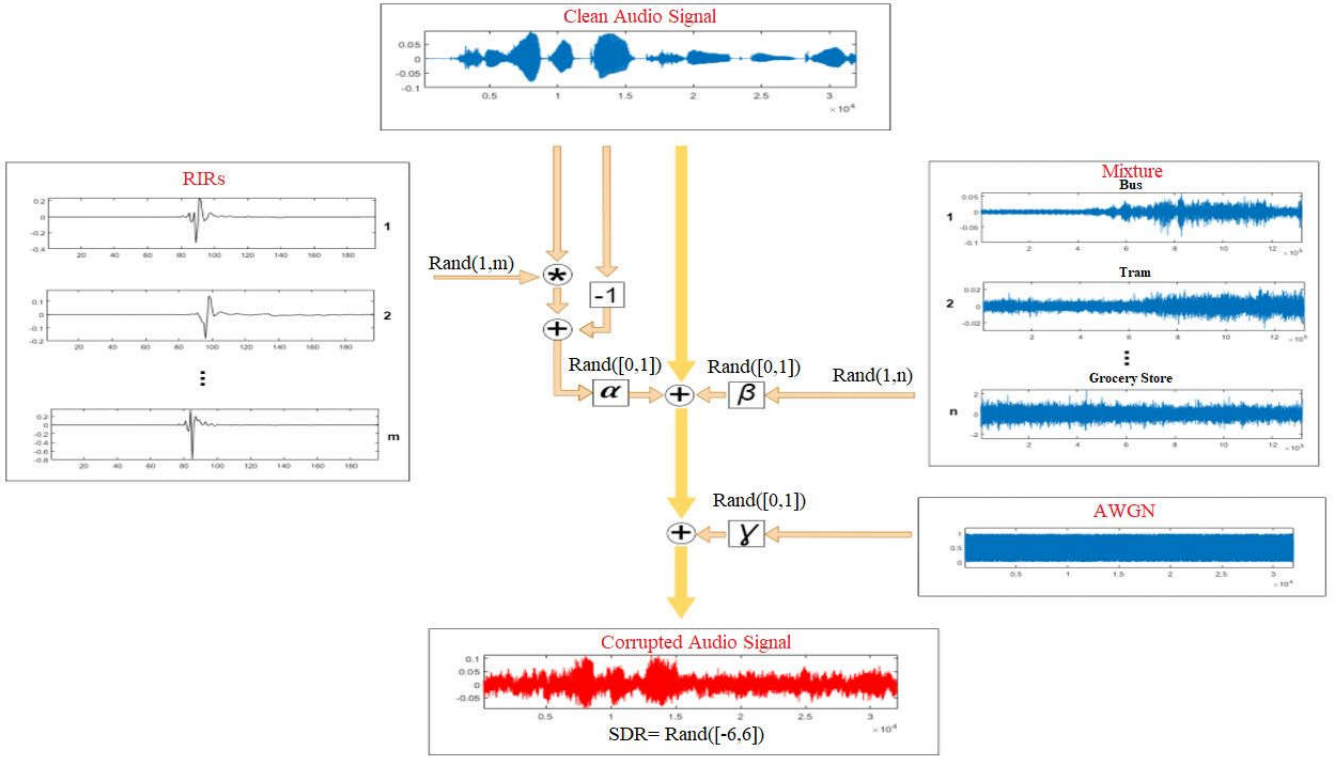

Figure 3: The illustration of a real-world audio dataset generation with a random choice of artifacts and their severities.

For each (32000 samples long) audio sample of both datasets, first, each corruption model is either randomly selected or discarded. At least one artifact type should be selected to enforce the generation of the corrupted audio. For the sake of clarity, we assume that all artifact types are selected for the following explanation. In the first stage, a binaural room impulse response (BRIR) from a set of 35 different BRIRs is convolved with the clean audio and the output scaled with a random parameter, $\alpha \sim U(0,1)$. Then, one of the background mixture models is randomly selected among 15 different audio sources: *beach, bus, cafe, car, city center, forest path, grocery store, home, library, metro, office, park, residential area, train, tram* and then scaled by another random parameter, $\beta \sim U(0,1)$. It is added to the corrupted audio in the first (reverberation) stage. Finally, 32000 samples of White Gaussian Noise (WGN) are generated and scaled with the third random parameter, $\gamma \sim U(0,1)$, and similarly, added to the corrupted audio in the second (reverberation + background mixture) stage. At the end of the third stage, the audio is, therefore, corrupted by all artifact types with random weights to yield a random SDR level in the range of -6 to +6 dB. If the final SDR level is not within this range, the same process is repeated with new (random) weights until the SDR level falls into this range. As a result, all clean audio segments with 32000 samples are corrupted by one (at least), two, or all artifact types with a final SDR level in the range of -6 to +6 dB.

### B. Experimental Setup

We used the same network architecture and training parameter settings for audio restoration experiments over TIMIT-RAR (speech) and GTZAN-RAR (non-speech) datasets. For the generator (G) of Op-GAN, a 10-layer U-Net configuration is used with 5 1-D self-operational layers and 5 upsampling (by 2) and self-operational layers with residual connections (instead of transposed convolution layers). This modification in the generator structure is proposed to avoid the creation of artifacts at the output of Op-GAN due to deconvolution. The kernel sizes are all set as 5. The stride is set as 2 for all operational layers. For the encoder side, the resulting dimensions of the feature maps per layer are 16000x16, 8000x32, 4000x64, 2000x128, and 1000x128.

In the decoder stage, the reverse upsampling and operational layers are used with generative neurons the number of which is doubled in each next layer. The Discriminator model consists of 6 self-operational layers with a kernel size of 4. The strides for layers are set as 2, 2, 2, 2, 1, and 2, respectively. As a loss function in the discriminator, mean squared error (MSE) is computed between the discriminator output and label vectors. The architectures for the generators and discriminators are shown in Figure 4. For all experiments, we employ a training scheme with a maximum of 1000 BP iterations and a batch size of 8. The Adam optimizer with the initial learning rates of $10^{-3}$ and $2.10^{-3}$ is used for the generator and the discriminator, respectively. The loss weights $\lambda_1$ and $\lambda_2$ in Eq. (9) are set as 10 and 5, respectively. We implemented the proposed 1D Self ONN architectures using the FastONN library [27] based on Python and PyTorch.

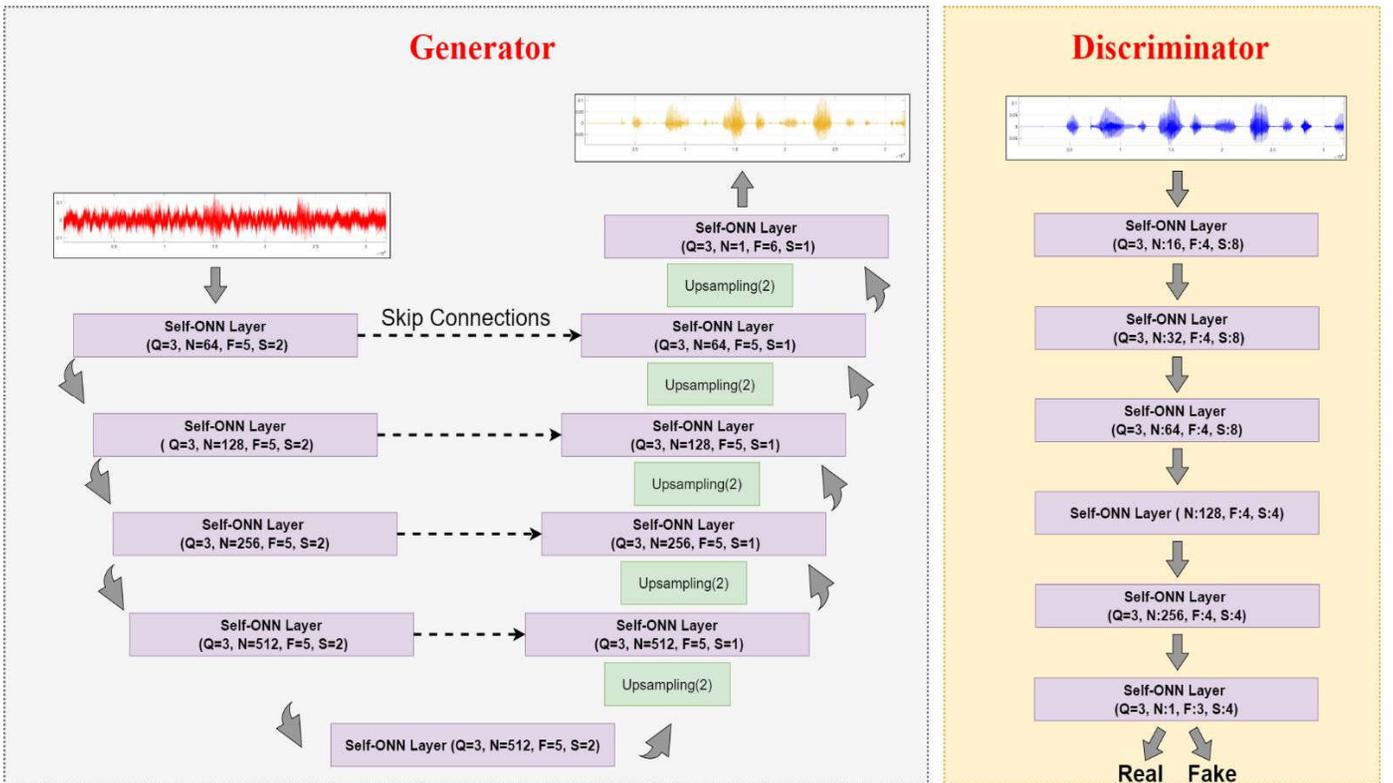

**Figure 4: The generator and discriminator architectures of the 1D Op-GANs.**

TABLE 2: OVERALL TEST PERFORMANCE OF OP-GAN OVER TIMIT-RAR AND GTZAN-RAR DATASETS.

|  |  | Mean SDR (dB) | Mean SegSNR (dB) | Mean FWSSNR (dB) |
|---|---|---|---|---|
| TIMIT-RAR | Corrupted Audio | 0.96 | -2.93 | 6.59 |
|  | Wiener [2] | 5.22 | 0.41 | 8.48 |
|  | SEGAN [7] | 5.30 | 0.35 | 6.71 |
|  | **Op-GAN** | **8.18** | **3.17** | **8.81** |
| GTZAN-RAR | Corrupted Audio | 1.18 | 0.53 | 8.34 |
|  | Wiener [2] | 2.96 | 2.39 | 6.88 |
|  | SEGAN [7] | 3.61 | 2.96 | 8.08 |
|  | **Op-GAN** | **6.06** | **5.32** | **10.27** |

TABLE 3: SPEECH-ONLY TEST PERFORMANCE OF OP-GAN OVER TIMIT-RAR DATASET.

|  |  | Mean STOI (%) | Mean PESQ | Mean CSIG | Mean CBAK | Mean COVL |
|---|---|---|---|---|---|---|
| TIMIT-RAR | Corrupted Audio | 78.31 | 1.31 | 1.96 | 1.69 | 1.58 |
|  | Wiener [2] | 79.3 | 1.44 | 2.21 | 1.91 | 1.75 |
|  | SEGAN [7] | 76.3 | 1.25 | 2.02 | 1.93 | 1.58 |
|  | **Op-GAN** | **82.44** | **1.46** | **2.56** | **2.26** | **1.97** |

### C. Quantitative Evaluations

For quantitative evaluations of the restored audio signals by the proposed and competing methods, the signal-to-distortion ratio (SDR), segmental SNR (SegSNR) [48], and frequency-weighted segmental SNR (FWSSNR) [49] metrics are used. Additionally, for the quantitative evaluation of the speech quality (for the TIMIT-RAR dataset), we also computed the short-time objective intelligibility (STOI), the perceptual evaluation of speech quality (PESQ), and other derived metrics such as CSIG, CBAK, and COVL using the implementation in [14]. CSIG predicts the mean opinion score (MOS) of the signal distortion, CBAK predicts the MOS of the background noise interferences, and COVL predicts the MOS of the overall speech quality.

TABLE 2 presents the performance of the proposed Op-GAN model evaluated using SDR, SegSNR, and FWSSNR metrics TIMIT-RAR and GTZAN-RAR datasets. For each metric, the mean values are computed over all (1453 and 1660, respectively) independent test samples. As a baseline from a well-known statistical reference method, the performance metrics for the Wiener filter based on a priori SNR estimation are computed for the same dataset [2].

In the literature, there are many GAN-based speech enhancement approaches, i.e., [7], [9], [10], [12], [14], and [43]. However, most of them have been specifically designed for only denoising or dereverberation, moreover assuming only one or a few degradation sources with pre-specified SNR levels. Consequently, one can expect that their restoration performance for real-world audio signals will be poor. To validate this fact and for comparative evaluations, we chose the waveform-based (1D) speech enhancement GAN (SEGAN) [7] which has a complex and deep generator (G) network with 22-layer 1D CNNs in an encoder-decoder architecture. Restorations by SEGAN over the two benchmark datasets are evaluated using its publicly shared trained weights [53].

The quantitative performance metrics are presented in TABLE 2. The proposed Op-GAN model achieved unprecedented mean SDR improvements, about 7.2 dB and 4.9 dB from the audio restoration experiments over the test sets of the TIMIT-RAR and GTZAN-RAR datasets, respectively. Significant performance improvements can also be observed for the SegSNR and FWSSNR metrics. On the other hand, the performance improvements of both competing methods are limited or even none in some cases, which is an expected outcome based on our earlier discussion. Overall, the Op-GAN model has a significant performance gap, and distortion suppression capability compared to the Wiener filter and the SEGAN model.

In TABLE 3, the commonly applied objective speech-only performance metrics are computed for the proposed Op-GAN model and the two competing methods over the test set of the TIMIT-RAR. The proposed Op-GAN model achieved the best mean restoration performances usually with a significant gap. On the other hand, the SEGAN model performs rather poorly due to the reasoning given earlier. The Op-GAN model further improved the audio quality in terms of noise and distortion levels (CSIG, CBAK, and COVL metrics) significantly better than the competitors.

Especially for the low SDR conditions as in the case TIMIT-RAR dataset, the first objective should be to suppress the noise and distortion levels sufficiently well while maintaining the intelligibility. Figure 5 shows the sample spectrograms for a sample corrupted audio at SDR=1.89 dB. For a visual comparison, Wiener-filtered and SEGAN-restored signal spectra are also presented. The Op-GAN achieved a significantly better restoration performance by suppressing the effects of different artifacts and boosting the output SDR level by around 14 dB, while the other two methods have SDRs of 8.23 dB and 8.82 dB, respectively.

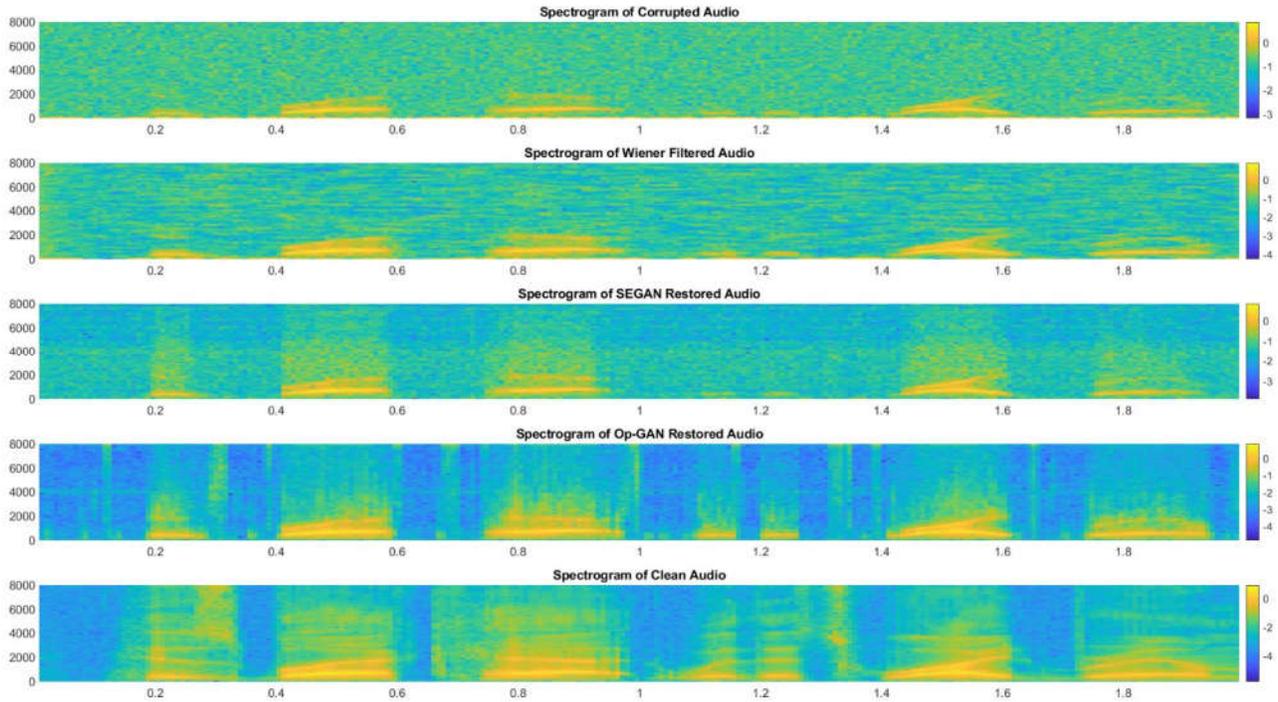

**Figure 5:** Sample spectrograms of clean (bottom) and corrupted (top) audio with SDR=1.89 dB. The restored audio signals by Wiener (2$^{nd}$ row), SEGAN (3$^{rd}$ row), and Op-GAN (4$^{th}$ row) have SDRs of 8.23 dB, 8.82 dB, and 14 dB, respectively.

### D. Qualitative Evaluation

Figure 6 to Figure 9 show four typical samples of the restored audio segments by the Op-GANs with the corresponding clean and corrupted counterparts from the TIMIT-RAR dataset. In these figures, each audio segment is corrupted with different corruption artifact\artifacts which are a background mixture, reverberation, AWGN, and blend of all artifacts, respectively. In Appendix A, we provided 10 more restoration results for both speech and non-speech data cases. Moreover, these audio segments are physically shared on the webpage [26] for an aural evaluation.

The first and foremost observation is that Op-GANs can achieve a crucial level of quality improvements regardless of the artifact type. In Figure 6, the audio signal is corrupted with a background mixture. This results in severe fluctuations and noise artifacts on the signal. The Op-GAN demonstrates an elegant level of restoration despite the severity of the corruption level (SDR = -3 dB).

Figure 7 shows a similar improvement over the reverberated signal. Due to the reverberation, the reflected sound masks the audio, and this results in lagging and overlapping on the clean signal. Op-GAN significantly reduces the lagging effect in non-speech parts and suppresses the overlapping almost completely. Overall, the SDR of the corrupted audio is improved from 3.63 dB to 6.22 dB.

In Figure 8, over the AWGN corrupted audio signal (SDR < 3dB), it can be seen that the Op-GAN suppressed the WGN almost completely and improved the SDR by approximately 10 dB. Finally, in Figure 9, a blend of all artifact types corrupted the audio segment where the Op-GAN successfully and significantly restored the audio and the SDR is improved from -5.14 dB to 6.68 dB. As seen from the zoomed first *silence* part of the signal, the Op-GAN does not generate any artifacts in this case.

Figure 10 and Figure 11 show two samples of the Op-GAN restored audio segments from the GTZAN-RAR dataset. In these figures, each audio segment is severely corrupted (input SDRs -1.02 and -1.65 dB) with a blend of all artifacts. In this case, the Op-GAN successfully and significantly restored the audio samples (notice the comparison for the zoomed last part of the record in Figure 11) and improved the SDRs up to 10.12 dB and 9.21 dB, respectively.

### E. Computational Complexity

For computational complexity analysis, the total number of parameters (PARs), and inference time (to restore an audio segment) for each network configuration are computed. The detailed formulations of the PARs calculations for Self-ONNs can be found in [27]. All the experiments were carried out on a 2.2 GHz Intel Core i7 with 16 GB of RAM and NVIDIA GeForce RTX 3080 graphic cards. For the implementation of the Op-GANs, Python with PyTorch library is used. Both the training and testing phases of the classifier are processed on the GPU cores. Note the inference time of the Op-GAN is computed for a single CPU. As shown in TABLE 4, for a single CPU implementation specifically, the required amount of time to restore a 1-second audio segment takes about 37.5 msec. This indicates more than 25 times the real-time speed on the test computer and therefore, the proposed audio restoration has the potential to be implemented on mobile, low-power devices in real-time. Especially, it can be a software plugin for mobile phones or other audio recorders. SEGAN, on the other hand, is not only more than 86% slower, it has 75 times more parameters than the Op-GAN.

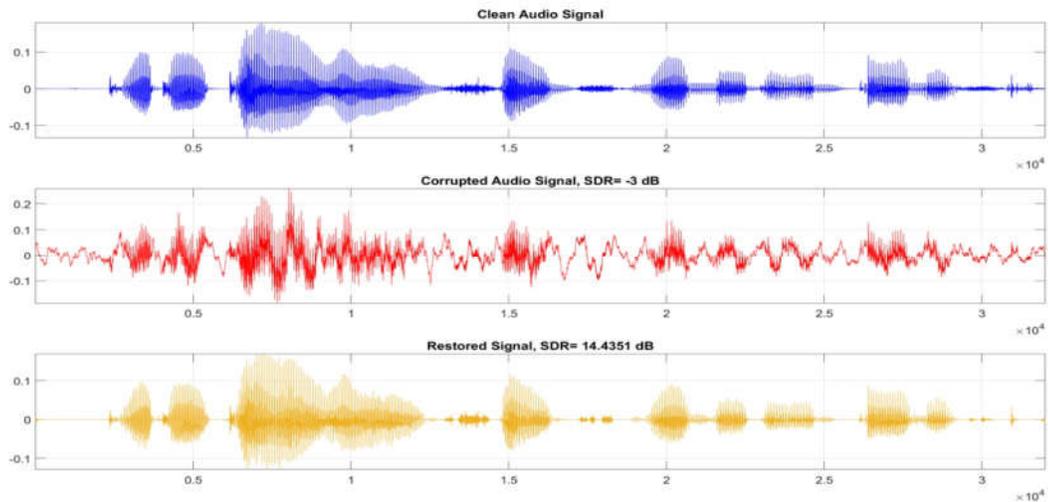

Figure 6: Sample test audio segment (from TIMIT-RAR) corrupted with the background mixture and the Op-GAN restored audio.

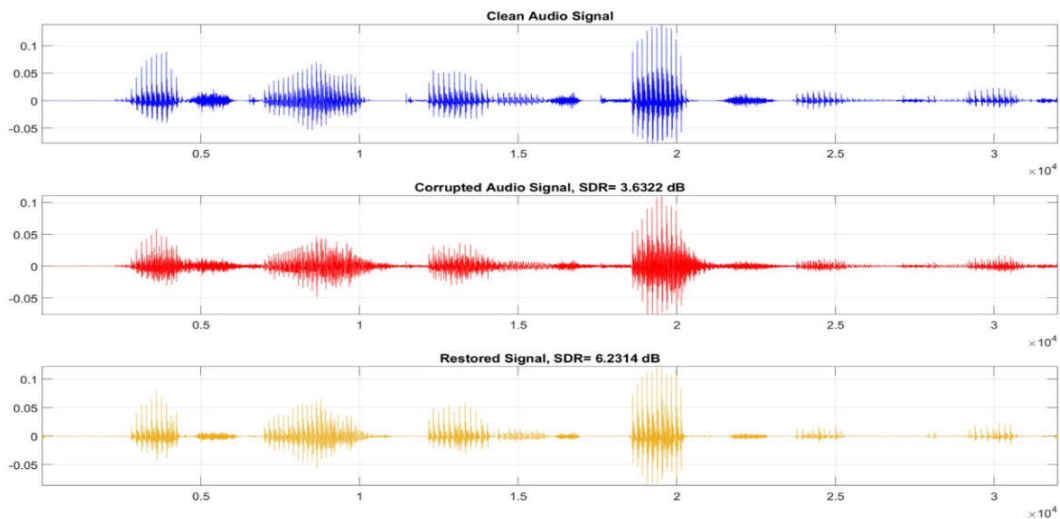

Figure 7: Sample test audio segment (from TIMIT-RAR) corrupted with the reverberation and the Op-GAN restored audio.

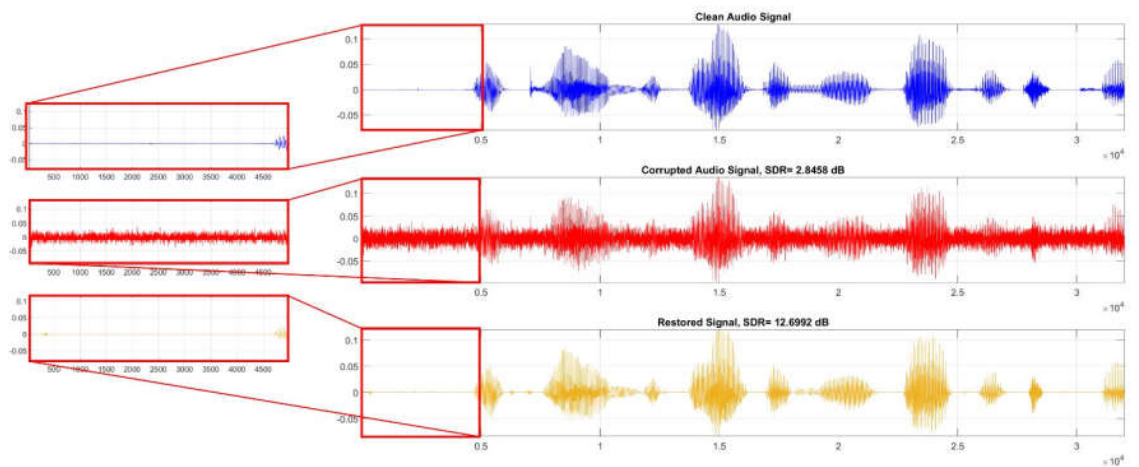

Figure 8: Sample test audio segment (from TIMIT-RAR) corrupted with the AWGN noise and the corresponding Op-GAN restored audio.

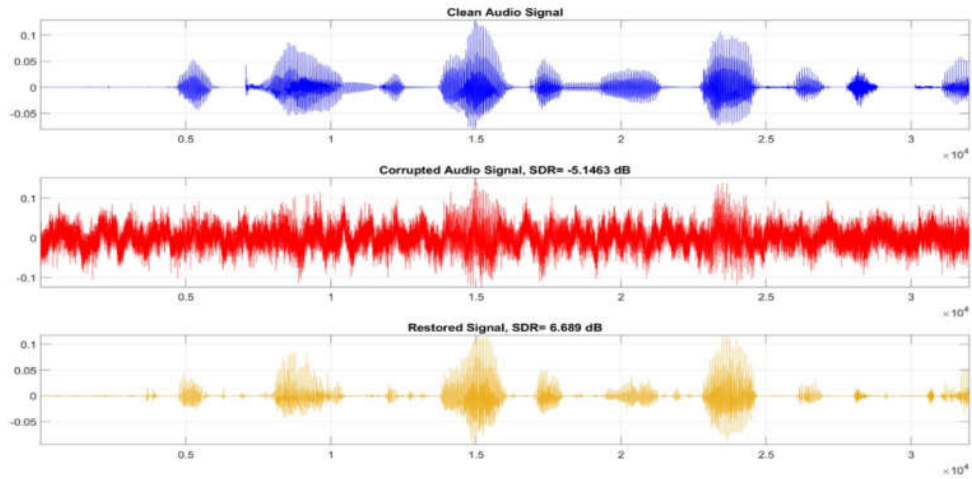

**Figure 9: Sample test audio segment (from TIMIT-RAR) corrupted with the blend of all artifacts and the Op-GAN restored audio.**

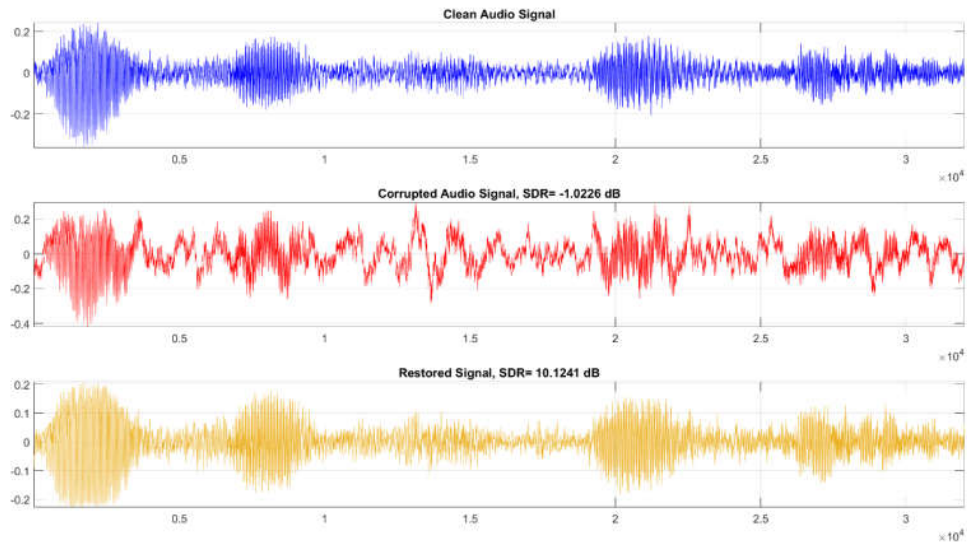

**Figure 10: Sample test audio segment (from GTZAN-RAR) corrupted with the blend of all artifacts and the Op-GAN restored audio.**

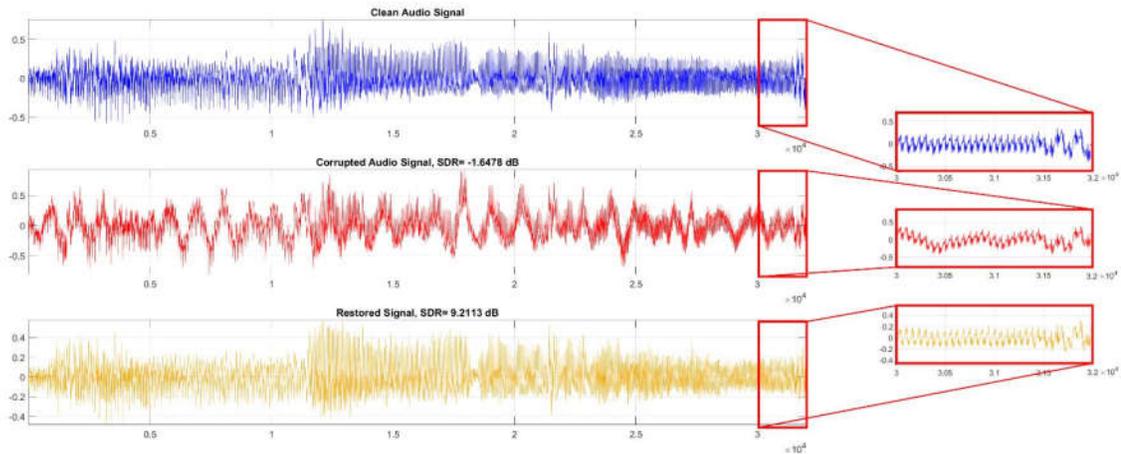

**Figure 11: Sample test audio segment (from GTZAN-RAR) corrupted with the blend of all artifacts and the corresponding Op-GAN restored audio.**

TABLE 4: COMPUTATIONAL COMPLEXITY OF THE OP-GANS.

| | PARs (K) | | | Inf. time (msec) per second |
|---|---|---|---|---|
| | G | D | Total | |
| 1D Op-GAN | 977 | 133 | 1110 | 37.5 |
| SEGAN [7] | 75,453 | 97,468 | 172,921 | 69.76 |

## IV. CONCLUSION

The major challenge of real-world audio restoration is that the acquired audio signals may severely be corrupted by a blend of artifacts such as reverberation, sensor noise, and background audio mixture. Each of these artifacts may vary in type, severity, and duration; therefore, the audio corruption usually shows a dynamic and non-stationary pattern along with the audio volume level. Despite numerous audio denoising and dereverberation works in the literature, no method has ever been proposed to address the restoration problem with such challenges. Therefore, in this study, to mimic real-world audio signals, we first compose the two benchmark datasets, a speech (TIMIT-RAR) and another non-speech (GTZAN-RAR) dataset. Using both datasets for training and evaluation, we propose a pioneer approach based on 1D Operational-GANs to perform a "blind" restoration for such real-world audio signals to achieve an elegant restoration performance and computational efficiency. Different from the prior works, the proposed *blind* restoration approach makes no prior assumptions about the artifact types and severity. As the baseline method, we proposed 1D Op-GAN for an end-to-end audio signal restoration system, and to further boost the performance a novel loss function is proposed for the generator to jointly utilize the temporal and spectral information of the audio signal. Once the Op-GAN model is trained over the clean and corrupted audio segments, the generator learns to suppress any artifact with any severity while preserving its characteristics and volume. The proposed 1D Op-GAN model is the most compact and shallowest model ever proposed among various GAN-based audio denoising and dereverberation models. Thus, it further achieves a minimal computational complexity that can lead to a real-time application even on low-cost, low-power devices and sensors. The optimized PyTorch code and the benchmark audio datasets are publicly shared in [26].

The quantitative and qualitative evaluations performed over an extensive set of audio recordings both benchmark datasets demonstrate that the quality of the severely corrupted audio signals can indeed be significantly improved for a wide range of SDRs. The proposed Op-GAN model achieved unprecedented SDR improvements, on average about 7.2 dB and 4.9 dB over the test sets of the TIMIT-RAR and GTZAN-RAR datasets, respectively. From the objective speech performance evaluations, the proposed model achieved overall 4% speech intelligibility (STOI) improvement and significantly less speech noise and distortion levels (CSIG, CBAK, and COVL metrics). These results further show that the novel Op-GAN model can 75directly be used to restore any audio clip corrupted with any blend of artifacts.. Finally, the performance of the blind audio restoration is the most evident especially on visual evaluations over the temporal plots and spectrograms of the clean and restored audio segments where all existing artifacts have significantly been suppressed even with such a compact generator model. This is not surprising considering the superiority of Self-ONNs in many challenging ML and CV tasks over the (deep) CNN models[15]-[32].

Over a large set of experiments, an important observation is that after the restoration, certain audio elements may be damaged or completely lost, especially when the corruption level is severe (i.e., SDR < 0 dB). We can foresee that such "restoration artifacts" can too be improved significantly by a second restoration pass of a new Op-GAN, which should now be designed and optimized accordingly since this is a completely different problem than real-world audio restoration. For this purpose, common quality metrics of certain audio types (e.g., speech, music, etc.) should be embedded into the adversarial loss function directly. Furthermore, spectral domain processing would play a crucial role to recover especially the high-frequency components of the audio that are altered/lost during the (first) restoration. This will be the main topic of our future research.

APPENDIX A

In this appendix, we provide 10 more audio restoration samples by the proposed Op-GAN model from both datasets as shown in the plots from Figure 12 to Figure 21. Additionally, all of these audio segments are shared on the webpage [26] for an aural evaluation.

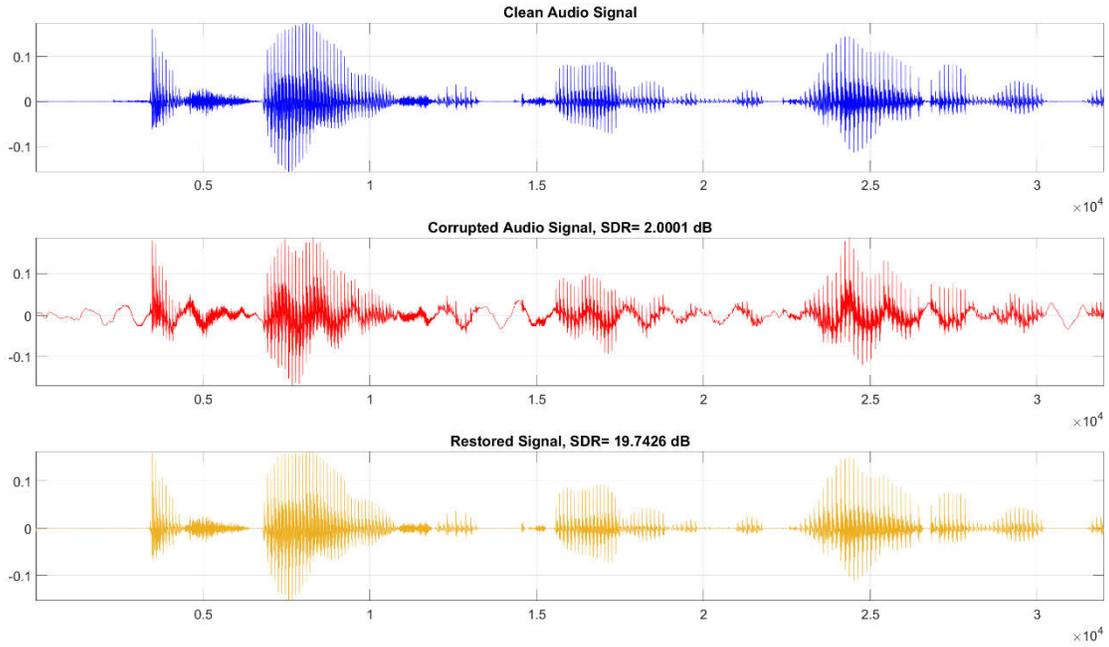

Figure 12: Sample test audio segment (from TIMIT-RAR) corrupted with the background mixture and the corresponding Op-GAN restored audio.

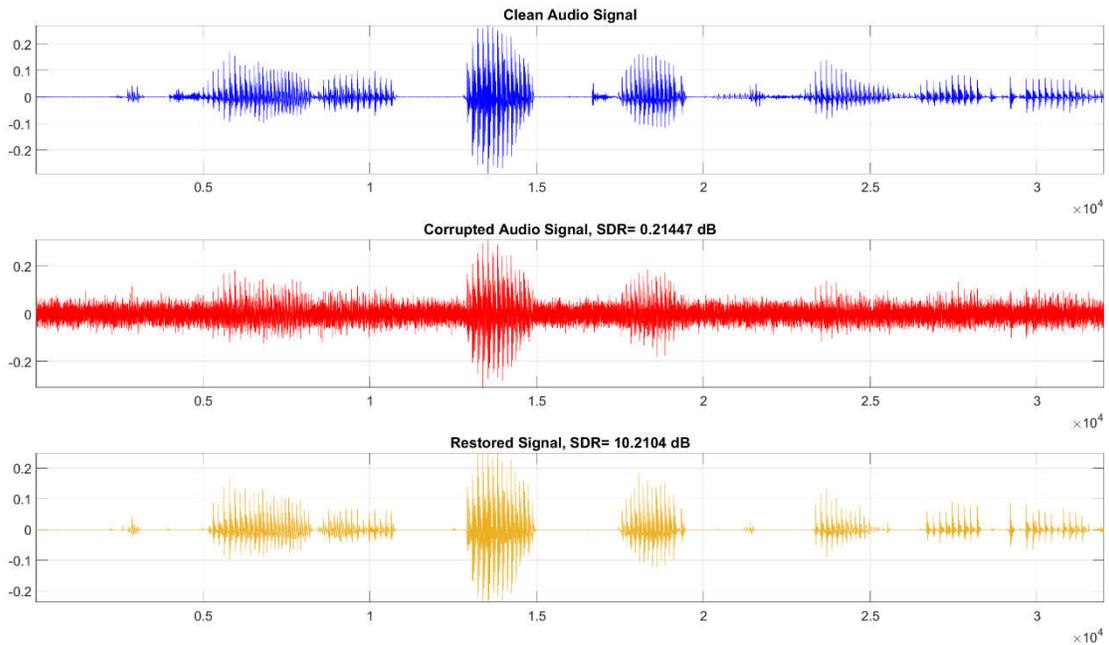

Figure 13: Sample test audio segment (from TIMIT-RAR) corrupted with the blend of all artifacts and the corresponding Op-GAN restored audio.
.

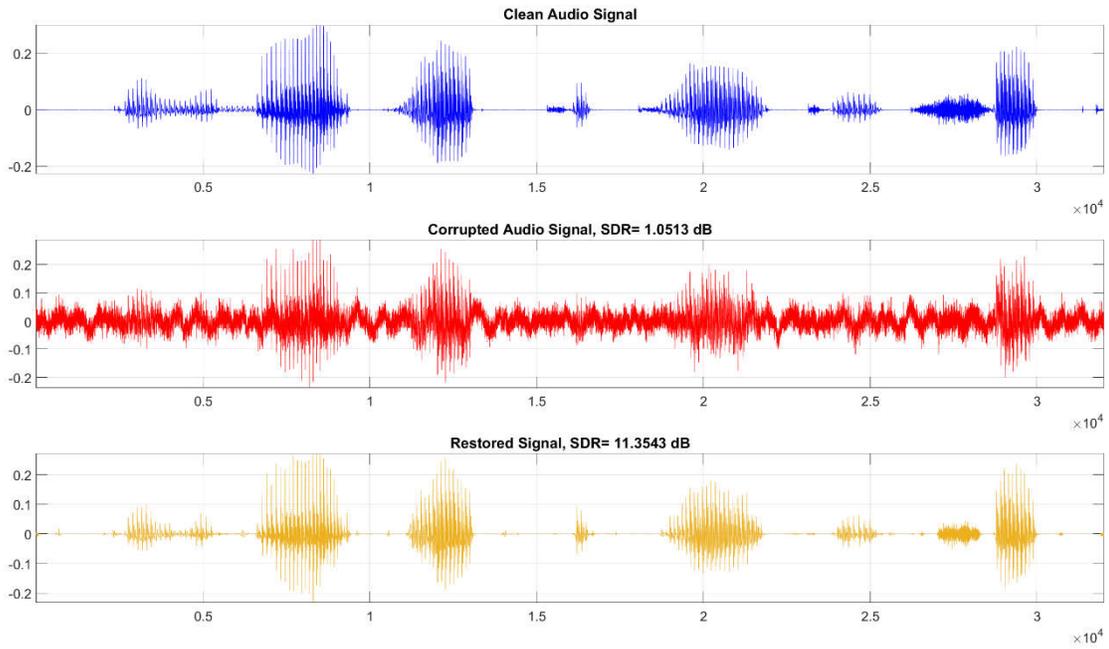

**Figure 14:** Sample test audio segment (from TIMIT-RAR) corrupted with the blend of all artifacts and the corresponding Op-GAN restored audio.

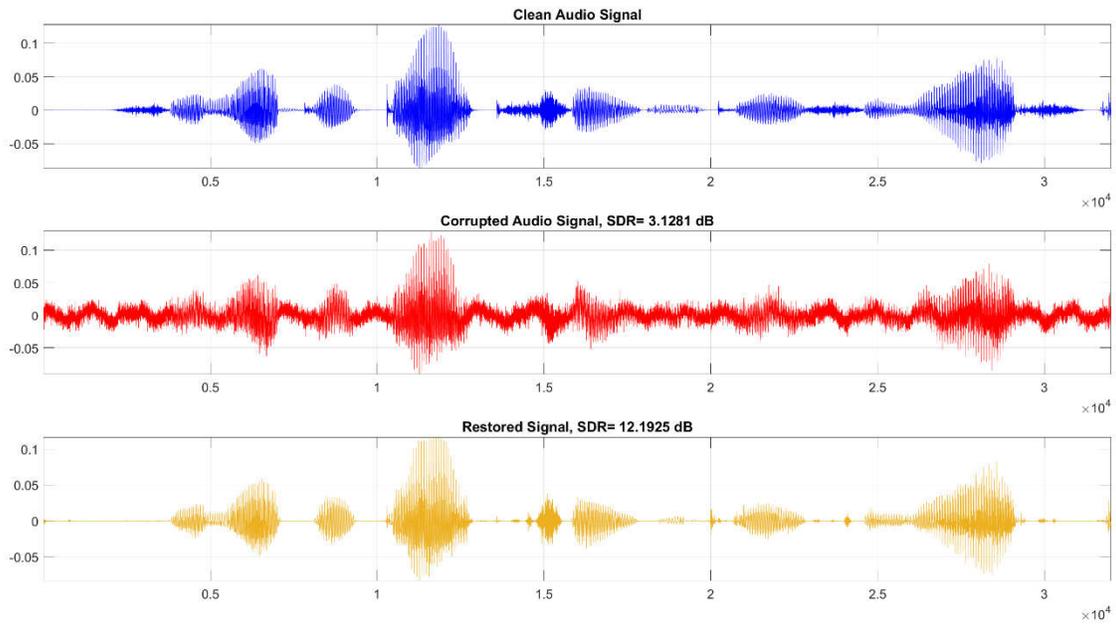

**Figure 15:** Sample test audio segment (from TIMIT-RAR) corrupted with the blend of all artifacts and the corresponding Op-GAN restored audio.

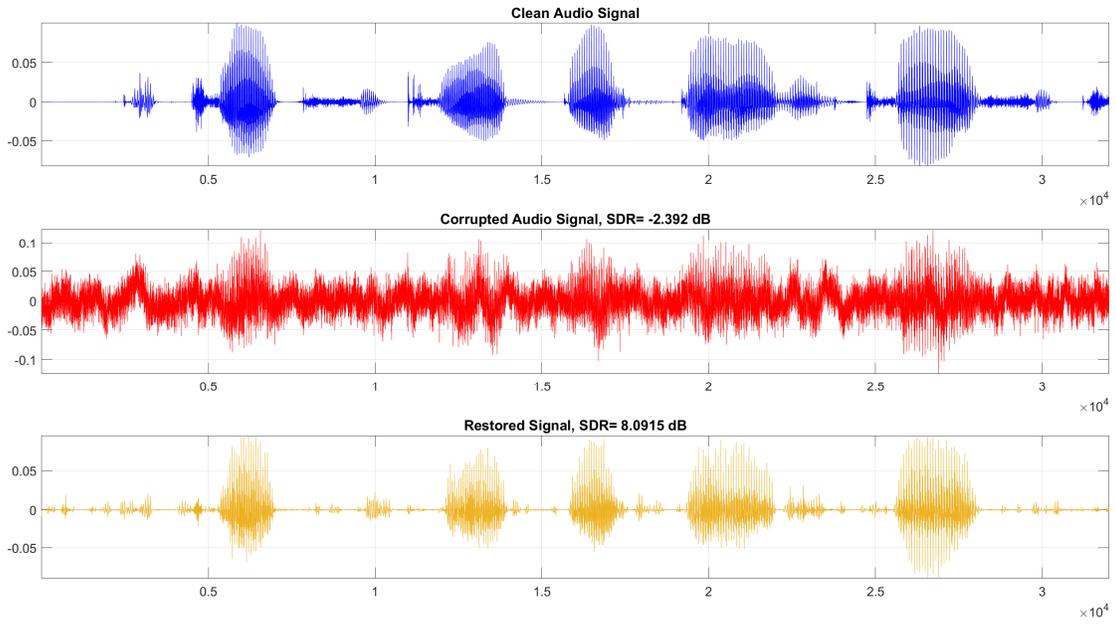

Figure 16: Sample test audio segment (from TIMIT-RAR) corrupted with the blend of all artifacts and the corresponding Op-GAN restored audio.

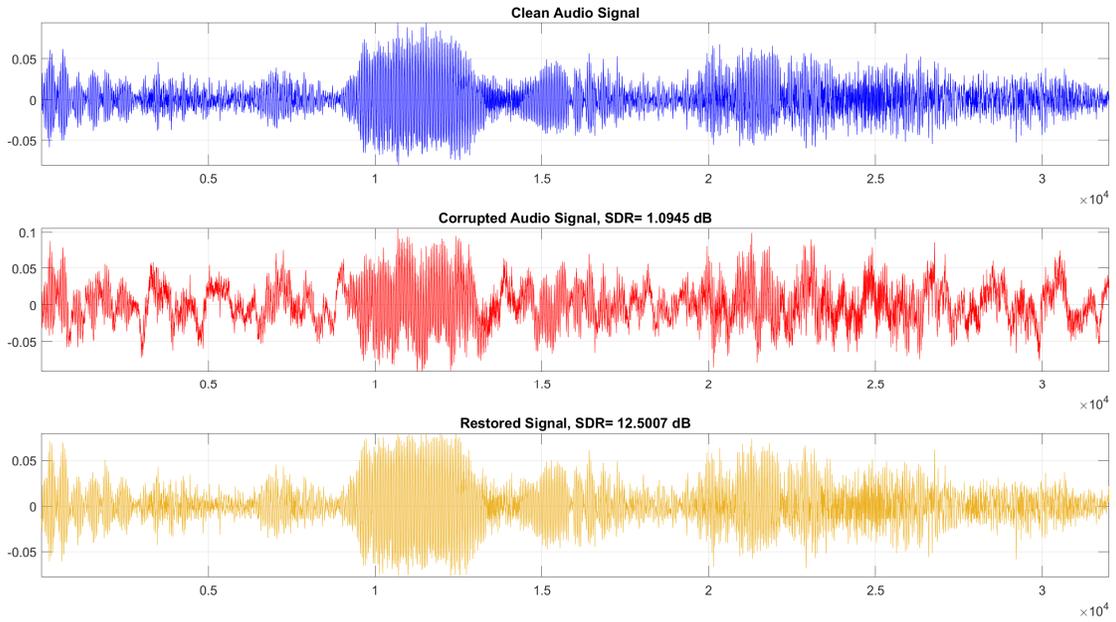

Figure 17: Sample test audio segment (from GTZAN-RAR) corrupted with the blend of all artifacts and the corresponding Op-GAN restored audio.

Blind Restoration of Real-World Audio by 1D Operational GANs

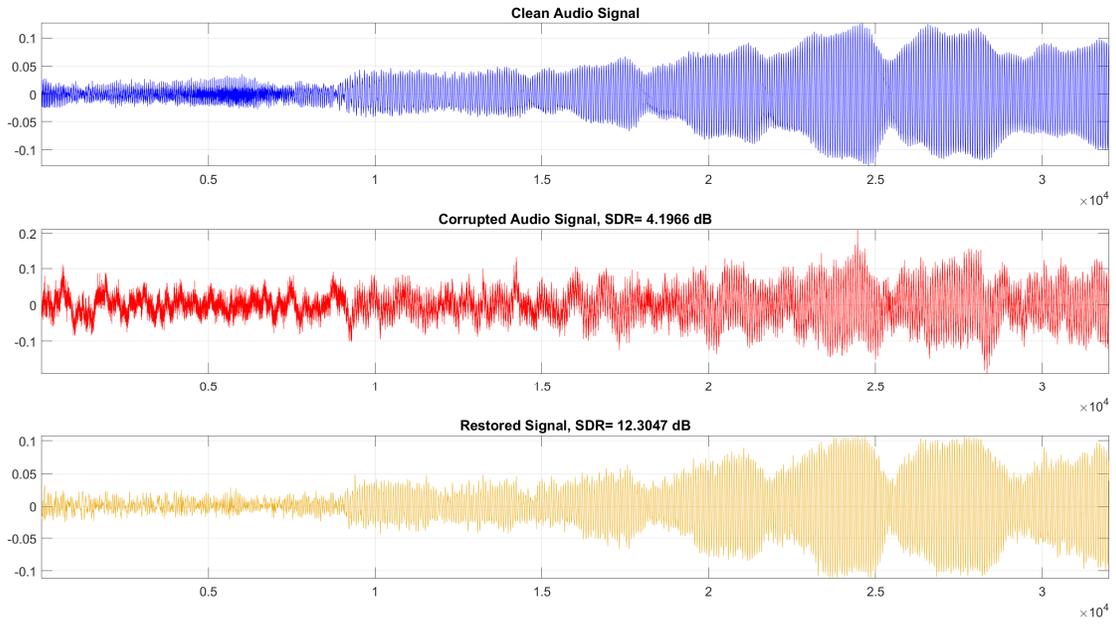

Figure 18: Sample test audio segment (from GTZAN-RAR) corrupted with the blend of all artifacts and the corresponding Op-GAN restored audio.

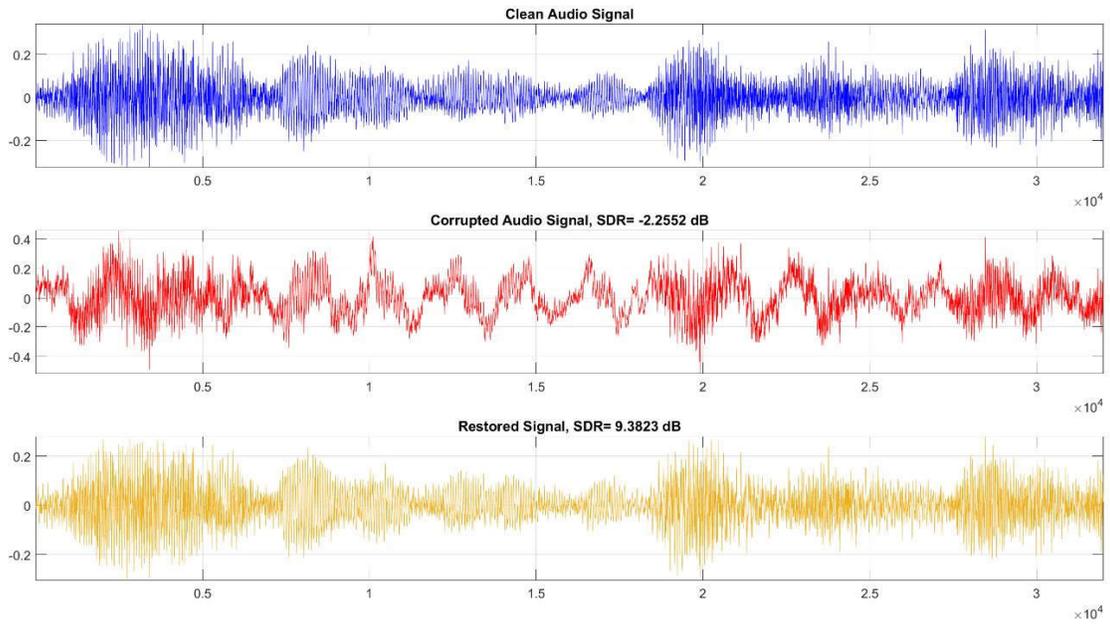

Figure 19: Sample test audio segment (from GTZAN-RAR) corrupted with the blend of all artifacts and the corresponding Op-GAN restored audio.

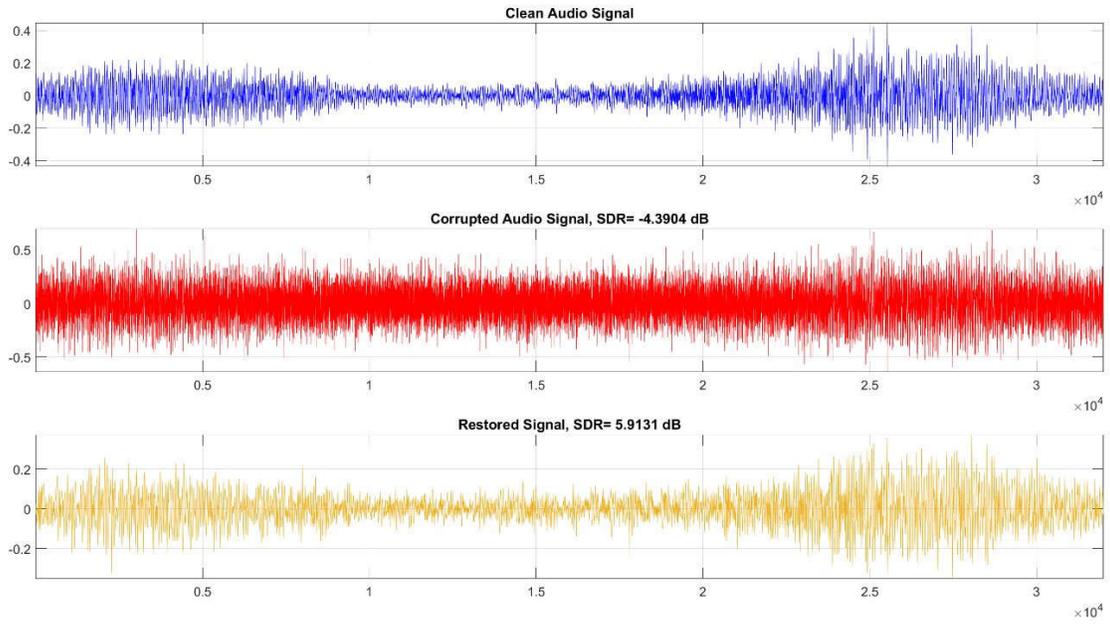

Figure 20: Sample test audio segment (from GTZAN-RAR) corrupted with the blend of all artifacts and the corresponding Op-GAN restored audio.

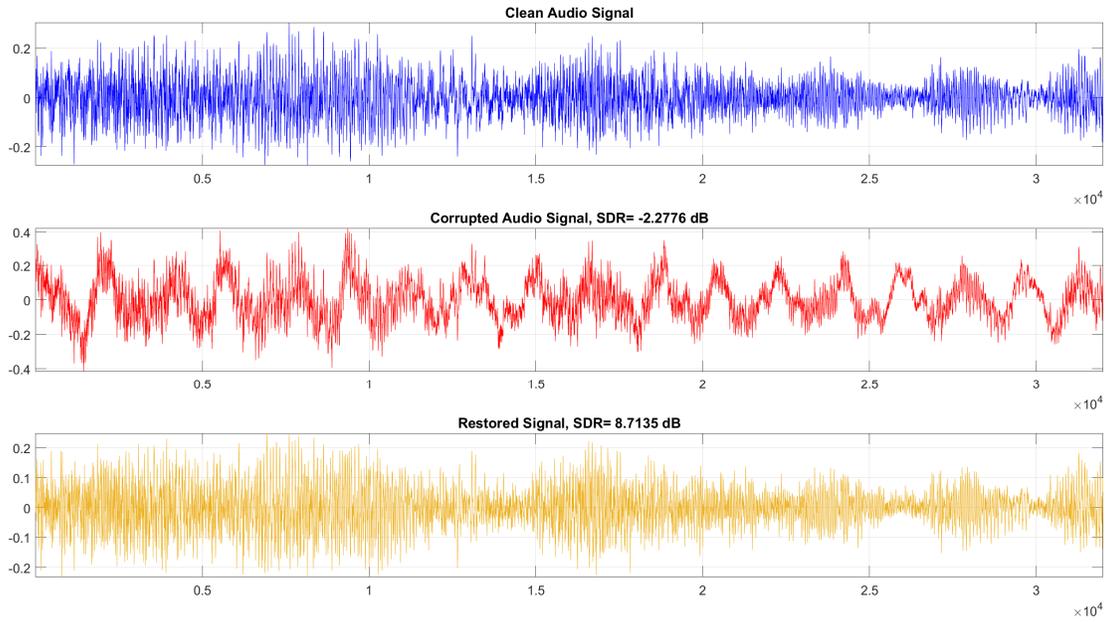

Figure 21: Sample test audio segment (from GTZAN-RAR) corrupted with the blend of all artifacts and the corresponding Op-GAN restored audio.